\documentclass{iau} 
\usepackage{graphicx}
\usepackage{natbib}
\usepackage{MnSymbol}
\usepackage{subcaption}
\usepackage{ulem}
\usepackage{xcolor}

\pubyear{2022}
\volume{370}
\jname{Winds of Stars and Exoplanets}
\editors{A.~A. Vidotto, L.~Fossati \& J.~Vink, eds.}

\title[Quantification of the environment of cool stars using numerical simulations] %% give here short title %%
{Quantification of the environment of cool stars using numerical simulations}

\author[Chebly et al.]   %% give here short author list %%
{J. J. Chebly $^1,^2$, J. D. Alvarado-Gómez$^1$ \and K.  Poppenhaeger$^1,^2$}

\affiliation{$^1$Leibniz Institute for Astrophysics, An der Sternwarte 16, 14482, Potsdam, Germany \\ [\affilskip]
$^2$Institute of Physics and Astronomy, University of Potsdam, Potsdam-Golm, 14476, Germany}

\begin{document}

\maketitle

\begin{abstract}

Stars interact with their planets through gravitation, radiation, and magnetic fields. Although magnetic activity decreases with time, reducing associated high-energy (e.g., coronal XUV emission, flares), stellar winds persist throughout the entire evolution of the system. Their cumulative effect will be dominant for both the star and for possible orbiting exoplanets, affecting in this way the expected habitability conditions. However, observations of stellar winds in low-mass main sequence stars are limited, which motivates the usage of models as a pathway to explore how these winds look like and how they behave. Here we present the results from a grid of 3D state-of-the-art stellar wind models for cool stars (spectral types F to M). We explore the role played by the different stellar properties (mass, radius, rotation, magnetic field) on the characteristics of the resulting magnetized winds (mass and angular momentum losses, terminal speeds, wind topology) and isolate the most important dependencies between the parameters involved. These results will be used to establish scaling laws that will complement the lack of stellar wind observational constraints.

%% add here a maximum of 10 keywords, to be taken from the file <Keywords.txt>
\end{abstract}

\firstsection % if your document starts with a section,

\section{Introduction}
The winds from cool, low-mass main sequence stars are weak and cannot be observed directly, except in the case of our own Sun. This means that the nature and behavior of these winds are not well understood.
More knowledge about stellar winds is needed because they carry a significant amount of angular momentum that affects the rotational evolution of the star (Johnstone et al. \citeyear{2015A&A...577A..28J}). 
These magnetized winds also affect the atmospheric evolution of  planetary atmospheres (Kislyakova et al. \citeyear{2014Sci...346..981K}). 
Several techniques have been proposed to infer the presence of stellar winds in cool stars. Examples are the free emission at radio wavelengths \citep{2000GeoRL..27..501G} and searches for X-ray emission induced by charge exchange \citep{2002ApJ...578..503W} between ionised stellar winds and neutral interstellar hydrogen. Unfortunately, both methods resulted in non-detections, but provided important upper limits on the wind mass-loss rates of a handful of Sun-like stars. One of the indirect techniques is the Ly-$\alpha$ absorption method, which uses high-resolution UV spectra from the Hubble Space Telescope (HST). As winds propagate from the star, they collide with the Inter Stellar Medium (ISM), forming 'astrospheres' analogous to the 'heliosphere' surrounding the Sun \citep{Wood2004}. Signs of charge exchange occur when the ISM is neutral or at least partially neutral, leading to astrospheric absorption signature in the Ly-$\alpha$ line.
This method is the most successful for measuring the mass loss rate ($\rm \dot{M}$) of winds from solar-like stars, with nearly 20 measurements to date \citep{Wood2021}. The new measurements showed, that the coronal activity and spectral type are not enough to determine wind properties.
However, this method has not provided measurements for larger numbers of stars because it is only applicable to stars within 10-15 pc. Beyond this distance, ISM becomes fully ionized. Also, Ly-$\alpha$ lines observed from cool stars are always heavily contaminated by interstellar absorption.
The lack of observational data and associated limitations motivate the use of numerical simulations as a way to improve our understanding of stellar winds.
Models based on the Alfvén waves are more popular when simulating the stellar wind from stars other than the Sun \citep{2006ApJ...640L..75S}. These waves are considered to be a likely key mechanism for solar wind heating and acceleration \citep{vanderHolst2014}.
In this study, we use one of the most realistic physics-based solar models to quantify the environment of low-mass main-sequence stars with an outer convective envelope, i.e., late-F to M type. Our goal is to provide a method for estimating stellar wind quantities as other related parameters become available. This will also help to further constrain the habitable zone (HZs) of exoplanets around these types of stars.

\section{Stellar Wind Model and Selected Sample}
\vspace{-0.4cm}
\begin{figure}[hbt!]
\begin{center}
 \includegraphics[width=1\textwidth]{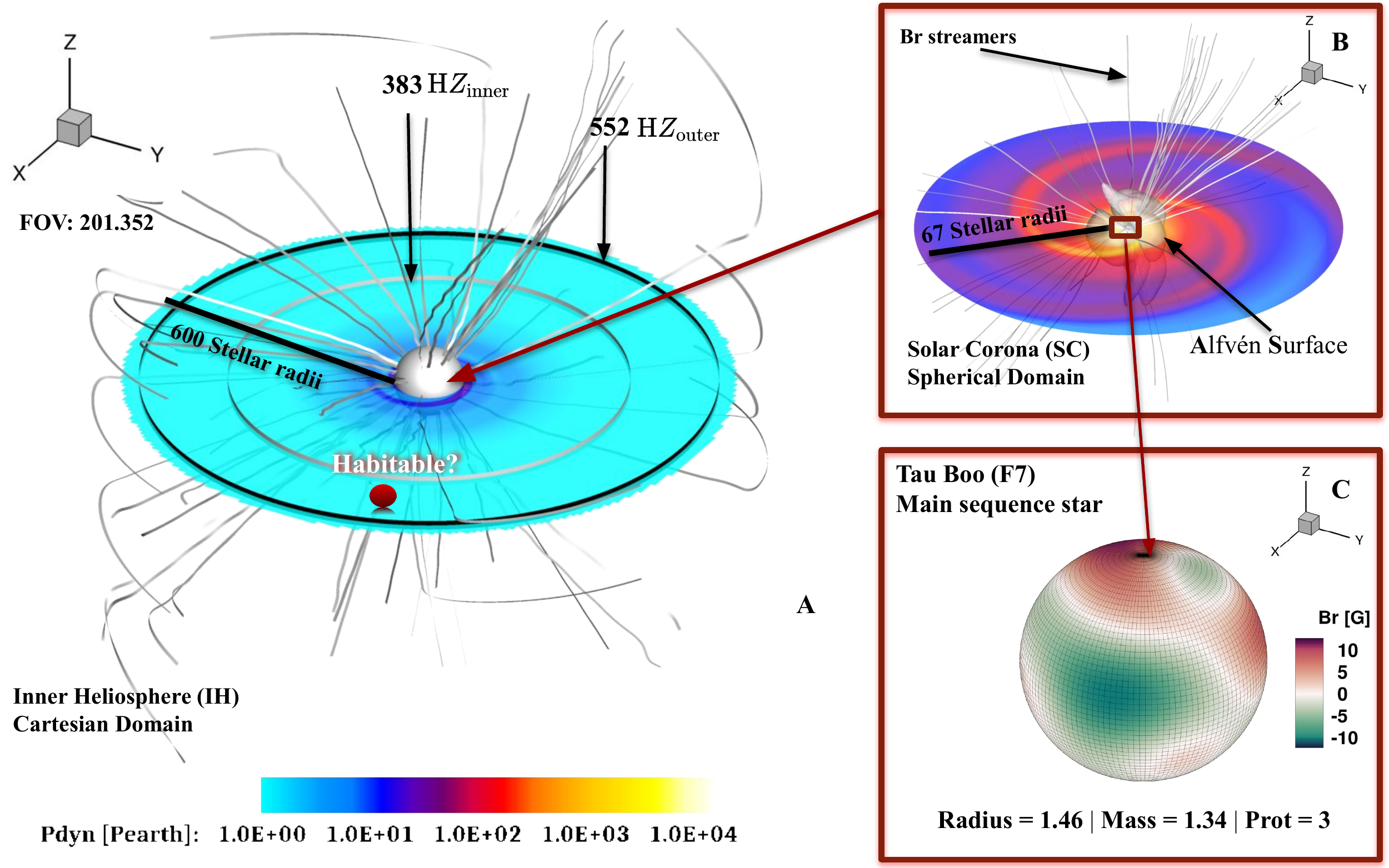} 
 \caption{Illustration of a 3D numerical magnetohydrodynamics simulation using the Alfv\'{e}n wave-solar model (AWSoM). Plots A and B show a steady-state solution of the stellar wind simulation of $\tau$ Boo (F7) and some stellar wind properties such as the dynamic pressure and the  Alfv\'{e}n surface ($S_{\rm A}$). In C, we show the magnetogram reconstructed from observed Zeeman Doppler Imaging maps used as the inner boundary of Solar Corona domain. The Inner Heliosphere (IH) is shown in plot A along with the limits of the habitable zone around $\tau$ Boo.} \label{fig1}
\end{center}
\end{figure}

We use the Alfv\'{e}n Wave Solar Model (AWSoM) to calculate different steady-state solutions for a grid of models. This model reproduces successfully the solar wind in a realistic manner (Oran et al. \citeyear{2013ApJ...778..176O}, Sachdeva et al. \citeyear{2019ApJ...887...83S}). Assuming that stellar winds from solar analogues are driven by the same process as the solar wind, we extend the model to stars other than the sun. 
AWSoM is part of the Space Weather Modelling Framework (SWMF), which integrates numerical models from the solar corona to the upper atmosphere into a high-performance coupled model \citep{Toth2012}.

In our simulation, we couple between two modules: the solar corona (SC), a spherical domain, and the inner heliosphere (IH), a Cartesian domain, as shown in Fig. \ref{fig1}, in the case of F, G, and K stars. Using the IH module was necessary for these stars in order to accommodate their large HZs. For M-dwarfs, the coupling was not necessary since the HZs is closer in.
The SC domain in our simulations ranges from 1.05 $R_{\filledstar}$ up to 67 $R_{\filledstar}$ for F, G, and K stars. 
For the M-dwarfs we require a sufficiently large SC domain (250~$R_{\filledstar}$) so  that the resulting $S_{\rm A}$ would be completely contained within it. Once the wind solution in SC reaches a steady state (converges), SC provides the plasma variables at the inner boundary of IH. Domain-overlap (from 62 $R_{\filledstar}$ to 67 $R_{\filledstar}$) is used in the coupling procedure between the two domains for F, G, and K type stars. The Inner Heliosphere component uses an ideal MHD approach and covers the range from 62 $R_{\filledstar}$ up to 600 R$_{\filledstar}$.

For our stellar wind models, the inner boundary conditions of AWSoM within the SC component (e.g. plasma density, temperature, Alfv\'{e}n wave pointing flux, and the Alfv\'{e}n wave correlation length), are kept identical to the values used to simulate the solar wind \citep{vanderHolst2014}.
In contrast, the magnetic field and stellar properties are modified for each star. The strength and geometry of the magnetic field is taken from from observed, reconstructed Zeeman Doppler Imaging (ZDI) maps. This is a tomographic imaging technique (Donati \& Brown \citeyear{1997A&A...326.1135D}; Kochukhov \& Piskunov \citeyear{2002A&A...388..868K}) that allows us to reconstruct the large-scale magnetic field (intensity and orientation) at the surface of the star from a series of circular polarization spectra \citep{2006MNRAS.370..629D}. Therefore, the simulations are more realistic compared to models that assume simplified/idealized field geometries.
We only retrieve the radial component of the magnetic field strength from the ZDI maps since solar wind models neglect the contribution from the other components.
We have also restricted the parameter space to stars with an outer convective envelope. The sample considered in this study consists of 21 stars (3F, 4G, 6K, 5M) located within 10 pc (taken from See et al. \citeyear{2019ApJ...876..118S}). Each star has its corresponding stellar rotation period ($P_{\rm rot}$), mass ($m_{\filledstar}$), and radius ($r_{\filledstar}$). The fastest star in our data set corresponds to an M6 star with $P_{\rm rot}$ = 0.71 d, and the slowest is a K3 star with a $P_{\rm rot}$ = 42.2 d. As for the magnetic field strength, it ranges from 5G to 2kG. 

\section{Numerical estimate of stellar wind properties}

By definition, the Alfv\'{e}n surface ($S_{\rm A}$) corresponds to the boundary between magnetically coupled outflows (MA $<$ 1) that do not carry angular momentum away from the star and the escaping stellar wind  (MA $>$ 1). In numerical models, the $S_{\rm A}$ is used to extract the mass loss rate ($\rm \dot{M}$), and the angular momentum loss rate ($\rm \dot{J}$) (e.g. Cohen et al. \citeyear{2010ApJ...723L..64C}; Garraffo et al. \citeyear{2015ApJ...813...40G}, Alvarado-Gómez et al. \citeyear{2016A&A...594A..95A}).
The angular momentum loss rate is obtained by integrating over the $S_{\rm A}$. Knowing the wind speed and density, the $\rm \dot{M}$ of a star can be calculated by integrating over a closed area enclosing the star beyond the $S_{\rm A}$. Note that the Alfv\'{e}n surface does not have a regular shape when visualized in 3D. This is due to several stellar parameters, such as the magnetic field distribution on the stellar surface, the wind velocity, and the density.
\vspace{-0.1cm}
\begin{equation}
\mathcal (\rm \dot{M}) = \rm {\rho}(\rm \textbf{u} \cdot dA)\label{eq1}
\end{equation}
\vspace{-0.5cm}
\begin{equation}
\mathcal (\rm \dot{J}) = \Omega\rho R^{2} \sin^{2} \theta (\textbf{u} \cdot dA)\label{eq2}
\end{equation}
\noindent Here $\rho$ represents the wind density, \textbf{u} is the wind speed vector and $dA$ is the vector surface element. $\rm \Omega$ is the angular frequency of the star and $\rm \theta$ is the angle between the lever arm and the rotation axis. $\Omega$ changes with the different stellar rotation $\Omega = 2\pi/\rm P_{\rm rot}$.

\subsection{$\rm \dot{M}$ and $\rm \dot{J}$ as a function of $B_{\rm R}$}
\vspace{-0.5cm}
\begin{figure}[ht]
  \subcaptionbox*{}[.54\linewidth]{%
    \includegraphics[width=\linewidth]{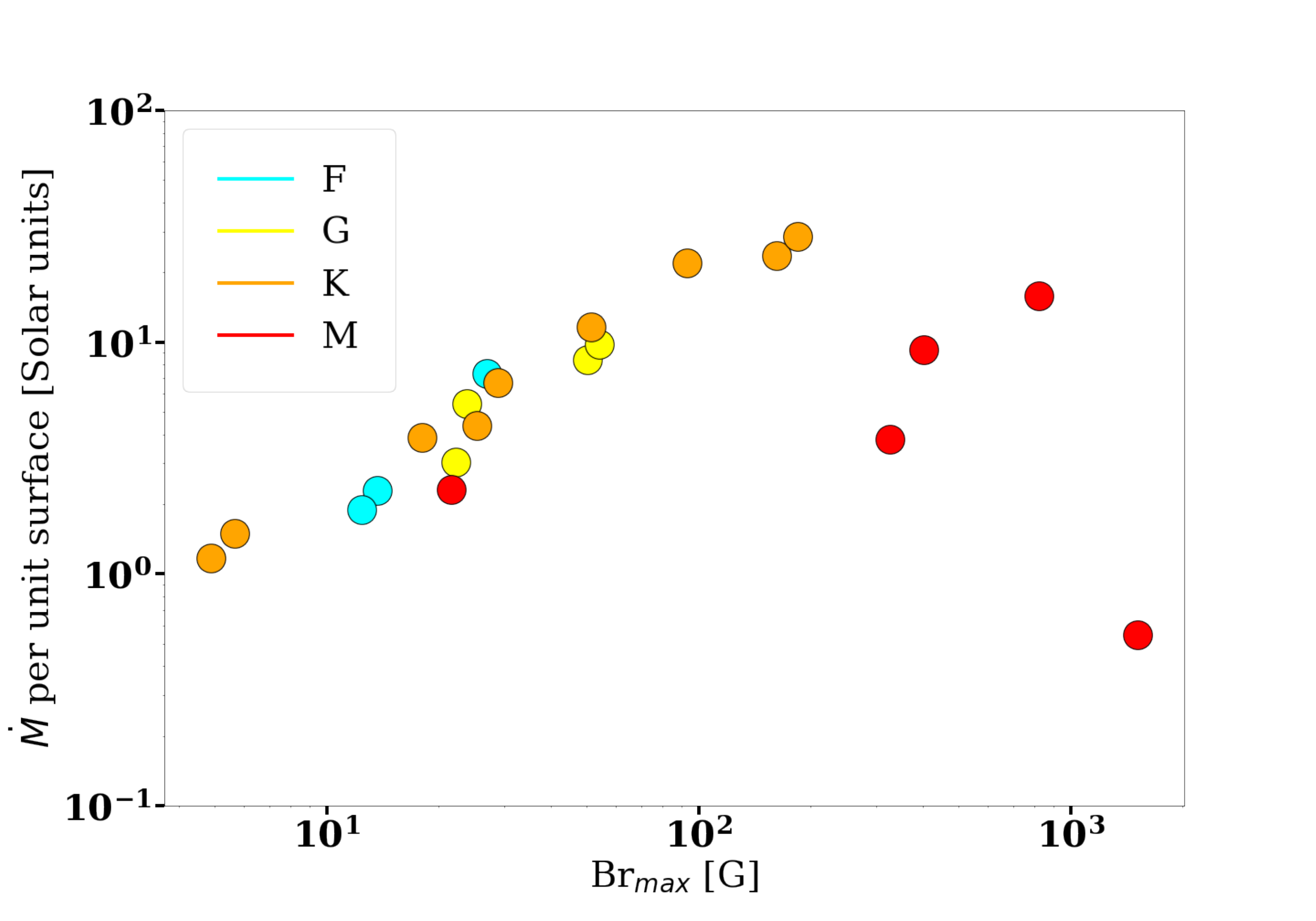}%
      \vspace{-0.5cm}
  }%
  \hfill
  \subcaptionbox*{}[.54\linewidth]{%
    \includegraphics[width=\linewidth]{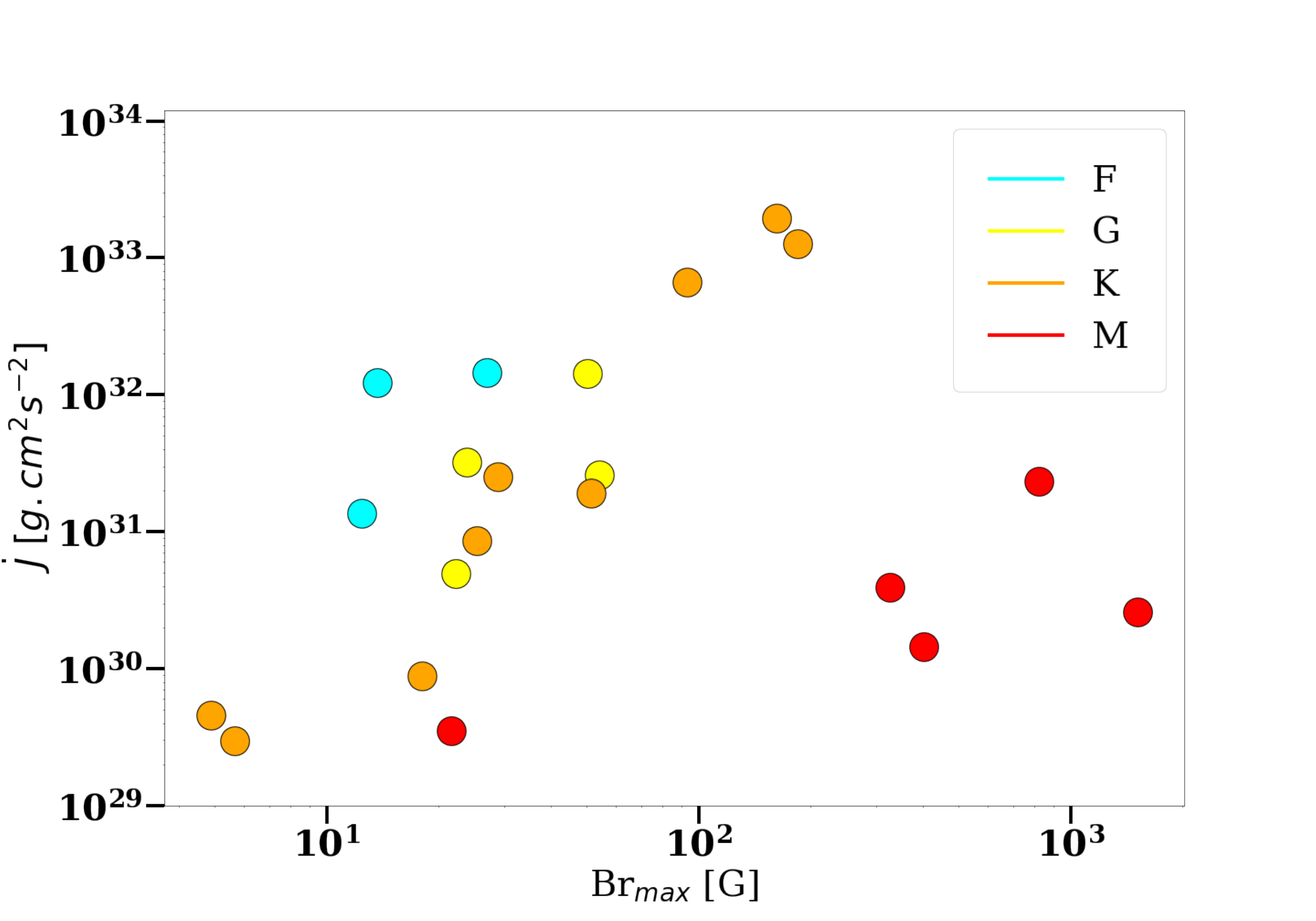}% 
     \vspace{-0.5cm}
  }
  \caption{Numerical estimate of $\rm \dot{M}$ and $\rm \dot{J}$ in function of the radial magnetic field ($B_{\rm R}$). The colors correspond to the different spectral types in our sample. The results shows that the loss rates are strongly dependent on the $B_{\rm R}$.}\label{fig2}
\end{figure}

In Fig. \ref{fig2}, we plot the the $\rm \dot{M}$ and $\rm \dot{J}$, as estimated using the method described previously, against $B_{\rm Rmax}$ for our sample of stars.
As expected, a stronger $B_{\rm R}$ produces stronger winds. This means either faster or denser winds. This interplay dictates $\rm \dot{M}$ and $\rm \dot{J}$ (cf. \ref{eq1}, \ref{eq2}), which will increase with increasing $B$ strength. We also note that we observe a strong dependence of $\rm \dot{M}$ and $\rm \dot{J}$ in stellar winds as a function of $B_{\rm Rmax}$, regardless of spectral type. 
In addition, we see on the left plot in figure \ref{fig2} the presence of an outlier in M-dwarfs for the strongest field $B_{\rm Rmax} = 1.5$ kG. One thing that might be causing this, is the way the simulation is being performed. We already pointed out that we had to push further out the SC boundary condition for the M-dwarfs. By doing this we reached a low grid resolution near the outer boundary which might have cause the $\rm \dot{M}$ to have a small value. Further testing will be performed to identify the possible cause of this outlier.

\subsection{$\rm \dot{M}$ as function of Rossby number}

The evolution of the magnetic field complexity was proved to be an important asset for explaining the bimodal distribution seen in the spin-down of stars that follows the Skumanich rotation evolution in young Open Clusters (OCs) \citep{2018ApJ...862...90G}. This complexity is a function of the Rossby number ($R_{\rm o}$) which is the ratio between the $P_{\rm rot}$ and the convective turnover time. The Rossby number is considered an important parameter for studying the effects of rotation on the dynamo action in stars with convective outer layers, being related to the dynamo number $D = R_{\rm o}^{-2}$ (Charbonnneau \citeyear{2020LRSP...17....4C}).
% \vspace{-0.2cm}
\begin{figure}[hbt!]
\begin{center}
  \includegraphics[width=1\textwidth]{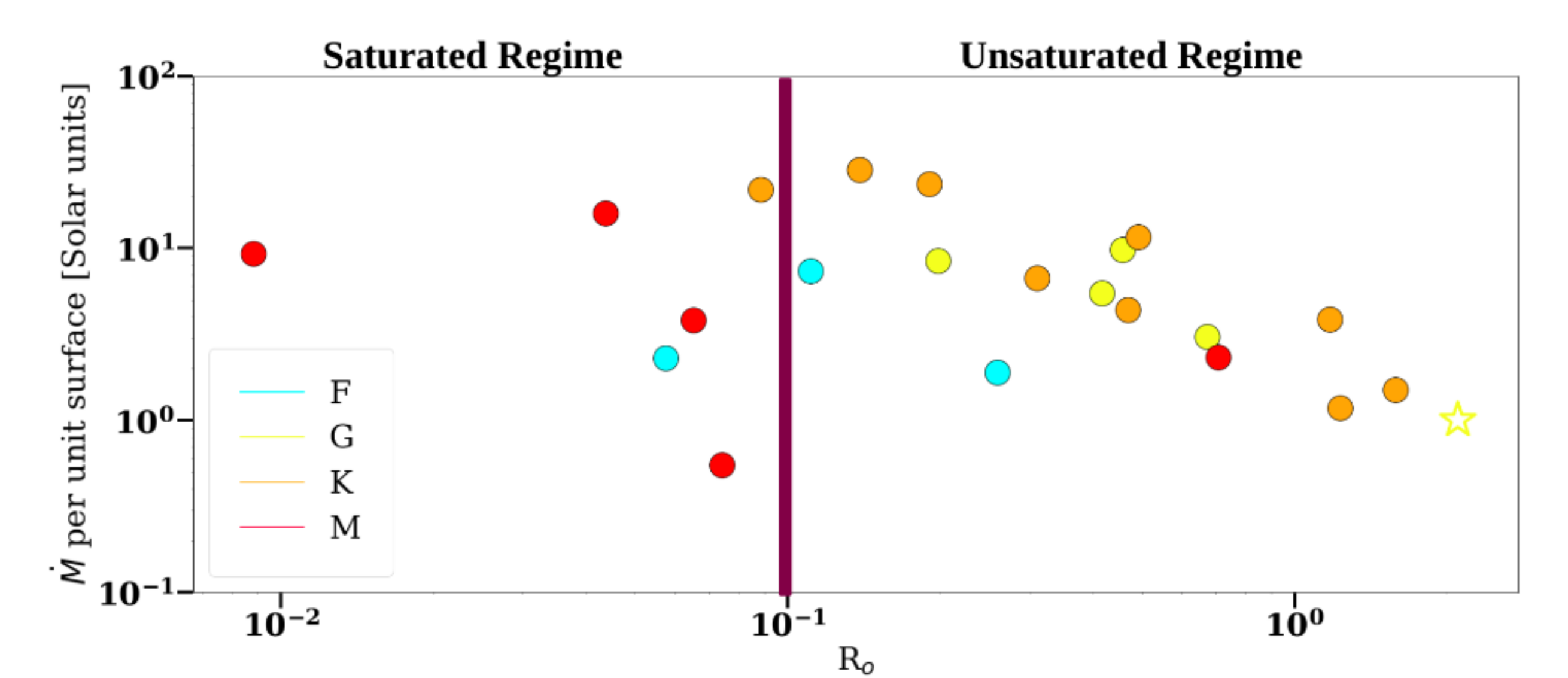} 
 \caption{3D MHD $\rm \dot{M}$ estimation in function of the Rossby number ($R_{\rm o}$). The colors correspond to the different spectral types in our sample and Solar value is represented by a star symbol. This plot highlights the importance of the field geometry in the magnetic field activity on the star surface.} \label{fig3}
\end{center}
\vspace{-0.1cm}
\end{figure}
We used the Rossby number as a magnetic field geometry parameter and explored how it changes with the $\rm \dot{M}$ loss rate of each star in our sample. We adapted empirical convective turnover times defined in \cite{Cranmer2011ApJ}, which depends on the effective temperature (Eq. \ref{eq4}). 
\vspace{-0.5cm}
\begin{equation}
\mathcal \rm \tau_{c} = 314.24  \exp {({\frac {- T_{\rm eff} } {1952.5 K} -  (\frac {T_{\rm eff}} {6250K})^{18}})}+ 0.002 \label{eq4}
\end{equation}
Our $\rm \dot{M}$ versus $R_{\rm o}$ (Fig. \ref{fig3}) shows a good representation of the trend for the normalized X-ray luminosity as a function of $R_{\rm o}$ in \cite{{2012LRSP....9....1R}} and Wright et al. (\citeyear{2018MNRAS.479.2351W}). The magnetic activity increases with decreasing $R_{\rm o}$ ($R_{\rm o}$ $\leq$ 0.1). At fast rotation ($R_{\rm o}$ = 0.1), the dynamo reaches a saturation level that cannot be exceeded even if the star rotates much faster.
Moreover, the simulated $\rm \dot{M}$ appears to have a maximum for stars with $R_{\rm o}$ $ < $ 0.1. This raises the question of whether stellar winds also reach a saturated regime.
We also point out that stars with different $R_{\rm o}$ can still have similar $\rm \dot{M}$. This is expected since the field complexity changes with rotation rate (Vidotto et al. \citeyear{{2014MNRAS.441.2361V}}a; Garraffo et al. \citeyear{{Garraffo2015}}; Réville et al. \citeyear{{2015ApJ...814...99R}}), which in turn influence the amount of mass loss.

\section{Discussion and Conclusion}

Understanding the behaviour and structure of cool stars' stellar winds and the habitability of its orbiting exoplanets requires detailed information on the strength and topology of the surface magnetic field. Analysis of high-resolution optical and NIR stellar intensity and polarisation spectra is currently the only approach that allows such information to be obtained in a systematic and direct manner. The measurement of quantities on the stellar surface is much simpler and does not suffer from problems with the ISM, which will always be a problem for the astrospheric and for the charge-exchange technique. 
Moreover, the number of cool stars with ZDI observations is growing ($\sim$ 200, Marsden et al. \citeyear{2014IAUS..302..138M}), so in principle estimates/models are possible for a larger number of stars that would allow to extract statistical properties about their winds. Quantifying these winds is becoming more crucial as the number of discovered exoplanets is increasing (currently 5178 confirmed exoplanets) and we enter a new era of detailed atmospheric characterization with JWST and E-ELTs. On the other hand, we are reaching the maximum number of systems that can be probed with the astrospheric technique due to the limitations mentioned earlier.
Three-dimensional MHD models that use ZDI maps as inner boundary conditions for the magnetic field strength and geometry, will help us obtain a more realistic estimate of the stellar wind parameters (e.g. dynamic pressure, terminal velocity, sub/super-Alfv\'enic conditions). Hence, we will be filling the missing observations and making progress in constraining the habitable conditions around low-mass main sequence stars.

\bibliographystyle{aasjournal}
\bibliography{macro_IAUS370.bib}

\begin{thebibliography}{}
\expandafter\ifx\csname natexlab\endcsname\relax\def\natexlab#1{#1}\fi
\providecommand{\url}[1]{\href{#1}{#1}}

\bibitem[{{Alvarado-G{\'o}mez} {et~al.}(2016){Alvarado-G{\'o}mez}, {Hussain},
  {Cohen}, {Drake}, {Garraffo}, {Grunhut}, \& {Gombosi}}]{2016A&A...594A..95A}
{Alvarado-G{\'o}mez}, J.~D., {Hussain}, G.~A.~J., {Cohen}, O., {et~al.} 2016,
  \aap, 594, A95

\bibitem[{{Charbonneau}(2020)}]{2020LRSP...17....4C}
{Charbonneau}, P. 2020, Living Reviews in Solar Physics, 17, 4

\bibitem[{{Cohen} {et~al.}(2010){Cohen}, {Drake}, {Kashyap}, {Sokolov}, \&
  {Gombosi}}]{2010ApJ...723L..64C}
{Cohen}, O., {Drake}, J.~J., {Kashyap}, V.~L., {Sokolov}, I.~V., \& {Gombosi},
  T.~I. 2010, \apjl, 723, L64

\bibitem[{{Cranmer} \& {Saar}(2011)}]{Cranmer2011ApJ}
{Cranmer}, S.~R., \& {Saar}, S.~H. 2011, \apj, 741, 54

\bibitem[{{Donati} \& {Brown}(1997)}]{1997A&A...326.1135D}
{Donati}, J.~F., \& {Brown}, S.~F. 1997, \aap, 326, 1135

\bibitem[{{Donati} {et~al.}(2006){Donati}, {Howarth}, {Jardine}, {Petit},
  {Catala}, {Landstreet}, {Bouret}, {Alecian}, {Barnes}, {Forveille},
  {Paletou}, \& {Manset}}]{2006MNRAS.370..629D}
{Donati}, J.~F., {Howarth}, I.~D., {Jardine}, M.~M., {et~al.} 2006, \mnras,
  370, 629

\bibitem[{{Gaidos} {et~al.}(2000){Gaidos}, {G{\"u}del}, \&
  {Blake}}]{2000GeoRL..27..501G}
{Gaidos}, E.~J., {G{\"u}del}, M., \& {Blake}, G.~A. 2000, \GeoRL, 27, 501

\bibitem[{{Garraffo} {et~al.}(2015{\natexlab{a}}){Garraffo}, {Drake}, \&
  {Cohen}}]{2015ApJ...813...40G}
{Garraffo}, C., {Drake}, J.~J., \& {Cohen}, O. 2015{\natexlab{a}}, \apj, 813,
  40

\bibitem[{{Garraffo} {et~al.}(2015{\natexlab{b}}){Garraffo}, {Drake}, \&
  {Cohen}}]{Garraffo2015}
---. 2015{\natexlab{b}}, \apjl, 807, L6

\bibitem[{{Garraffo} {et~al.}(2018){Garraffo}, {Drake}, {Dotter}, {Choi},
  {Burke}, {Moschou}, {Alvarado-G{\'o}mez}, {Kashyap}, \&
  {Cohen}}]{2018ApJ...862...90G}
{Garraffo}, C., {Drake}, J.~J., {Dotter}, A., {et~al.} 2018, \apj, 862, 90

\bibitem[{{Johnstone} {et~al.}(2015){Johnstone}, {G{\"u}del}, {Brott}, \&
  {L{\"u}ftinger}}]{2015A&A...577A..28J}
{Johnstone}, C.~P., {G{\"u}del}, M., {Brott}, I., \& {L{\"u}ftinger}, T. 2015,
  \aap, 577, A28

\bibitem[{{Kislyakova} {et~al.}(2014){Kislyakova}, {Holmstr{\"o}m}, {Lammer},
  {Odert}, \& {Khodachenko}}]{2014Sci...346..981K}
{Kislyakova}, K.~G., {Holmstr{\"o}m}, M., {Lammer}, H., {Odert}, P., \&
  {Khodachenko}, M.~L. 2014, Science, 346, 981

\bibitem[{{Kochukhov} \& {Piskunov}(2002)}]{2002A&A...388..868K}
{Kochukhov}, O., \& {Piskunov}, N. 2002, \aap, 388, 868

\bibitem[{{Marsden} {et~al.}(2014){Marsden}, {Petit}, {Jeffers}, {do
  Nascimento}, {Carter}, \& {Brown}}]{2014IAUS..302..138M}
{Marsden}, S., {Petit}, P., {Jeffers}, S., {et~al.} 2014, in Magnetic Fields
  throughout Stellar Evolution, ed. P.~{Petit}, M.~{Jardine}, \& H.~C.
  {Spruit}, Vol. 302, 138--141

\bibitem[{{Oran} {et~al.}(2013){Oran}, {van der Holst}, {Landi}, {Jin},
  {Sokolov}, \& {Gombosi}}]{2013ApJ...778..176O}
{Oran}, R., {van der Holst}, B., {Landi}, E., {et~al.} 2013, \apj, 778, 176

\bibitem[{{Reiners}(2012)}]{2012LRSP....9....1R}
{Reiners}, A. 2012, Living Reviews in Solar Physics, 9, 1

\bibitem[{{R{\'e}ville} {et~al.}(2015){R{\'e}ville}, {Brun}, {Strugarek},
  {Matt}, {Bouvier}, {Folsom}, \& {Petit}}]{2015ApJ...814...99R}
{R{\'e}ville}, V., {Brun}, A.~S., {Strugarek}, A., {et~al.} 2015, \apj, 814, 99

\bibitem[{{Sachdeva} {et~al.}(2019){Sachdeva}, {van der Holst}, {Manchester},
  {T{\'o}th}, {Chen}, {Lloveras}, {V{\'a}squez}, {Lamy}, {Wojak}, {Jackson},
  {Yu}, \& {Henney}}]{2019ApJ...887...83S}
{Sachdeva}, N., {van der Holst}, B., {Manchester}, W.~B., {et~al.} 2019, \apj,
  887, 83

\bibitem[{{See} {et~al.}(2019){See}, {Matt}, {Folsom}, {Boro Saikia}, {Donati},
  {Fares}, {Finley}, {H{\'e}brard}, {Jardine}, {Jeffers}, {Lehmann}, {Marsden},
  {Mengel}, {Morin}, {Petit}, {Vidotto}, {Waite}, \& {BCool
  Collaboration}}]{2019ApJ...876..118S}
{See}, V., {Matt}, S.~P., {Folsom}, C.~P., {et~al.} 2019, \apj, 876, 118

\bibitem[{{Suzuki}(2006)}]{2006ApJ...640L..75S}
{Suzuki}, T.~K. 2006, \apjl, 640, L75

\bibitem[{{T{\'o}th} {et~al.}(2012){T{\'o}th}, {van der Holst}, {Sokolov}, {De
  Zeeuw}, {Gombosi}, {Fang}, {Manchester}, {Meng}, {Najib}, {Powell}, {Stout},
  {Glocer}, {Ma}, \& {Opher}}]{Toth2012}
{T{\'o}th}, G., {van der Holst}, B., {Sokolov}, I.~V., {et~al.} 2012, Journal
  of Computational Physics, 231, 870

\bibitem[{{Van der Holst} {et~al.}(2014){Van der Holst}, {Sokolov}, {Meng},
  {Jin}, {Manchester}, {T{\'o}th}, \& {Gombosi}}]{vanderHolst2014}
{Van der Holst}, B., {Sokolov}, I.~V., {Meng}, X., {et~al.} 2014, \apj, 782, 81

\bibitem[{{Vidotto} {et~al.}(2014){Vidotto}, {Gregory}, {Jardine}, {Donati},
  {Petit}, {Morin}, {Folsom}, {Bouvier}, {Cameron}, {Hussain}, {Marsden},
  {Waite}, {Fares}, {Jeffers}, \& {do Nascimento}}]{2014MNRAS.441.2361V}
{Vidotto}, A.~A., {Gregory}, S.~G., {Jardine}, M., {et~al.} 2014, \mnras, 441,
  2361

\bibitem[{{Wargelin} \& {Drake}(2002)}]{2002ApJ...578..503W}
{Wargelin}, B.~J., \& {Drake}, J.~J. 2002, \apj, 578, 503

\bibitem[{{Wood}(2004)}]{Wood2004}
{Wood}, B.~E. 2004, Living Reviews in Solar Physics, 1, 2

\bibitem[{{Wood} {et~al.}(2021){Wood}, {M{\"u}ller}, {Redfield}, {Konow},
  {Vannier}, {Linsky}, {Youngblood}, {Vidotto}, {Jardine},
  {Alvarado-G{\'o}mez}, \& {Drake}}]{Wood2021}
{Wood}, B.~E., {M{\"u}ller}, H.-R., {Redfield}, S., {et~al.} 2021, \apj, 915,
  37

\bibitem[{{Wright} {et~al.}(2018){Wright}, {Newton}, {Williams}, {Drake}, \&
  {Yadav}}]{2018MNRAS.479.2351W}
{Wright}, N.~J., {Newton}, E.~R., {Williams}, P. K.~G., {Drake}, J.~J., \&
  {Yadav}, R.~K. 2018, \mnras, 479, 2351

\end{thebibliography}

\end{document}